\documentstyle[12pt,epsfig,wrapfig]{article}
\begin{document}
\newcommand{\bfk}{\mbox{\boldmath $k$}}
\begin{flushright}
INFNCA-TH9621 \\
DFTT 67/96 \\
hep-ph/9610387 \\
October 1996
\end{flushright}
\vspace{0.5truecm}
\renewcommand{\thefootnote}{\fnsymbol{footnote}}
\begin{center}
{\large \bf
Single Spin Asymmetries in Inclusive Hadron Production\footnote{
\,Talk delivered by F. Murgia at the XII International
Symposium on High Energy Spin Physics, Amsterdam, Sept. 10-14, 1996.}
\\ }
\vspace{5mm}
M. Anselmino$^1$, M. Boglione$^1$ and F. Murgia$^2$ \\
\vspace{5mm}
{\small\it (1) Dipartimento di Fisica Teorica, Universit\`a di Torino and \\
  INFN, Sezione di Torino, Via P. Giuria 1, 10125 Torino, Italy } \\
{\small\it (2) INFN, Sezione di Cagliari, Via A. Negri 18,
               09127 Cagliari, Italy}
\end{center}

\begin{center}
ABSTRACT

\vspace{5mm}
\begin{minipage}{130 mm}
\small
We consider a general formalism, previously developed to describe
single spin asymmetries (SSA) in inclusive particle production at
moderately large $p_T$ in hadron-hadron high-energy processes.
In this approach SSA can originate from non perturbative, universal,
quark distribution analyzing powers, provided the quark intrinsic motion
is taken into account. A fit of the available data on
$p^\uparrow\, p \to \pi\, X$ fixes simplified phenomenological
expressions of these quark analyzing
powers for the dominant (at large $x_F$) $u$ and $d$ quark contributions.
We use the results of the fit as input to provide a possible
scenario for SSA
in a number of interesting processes ({\it e.g.} $p\, p \to \gamma\, X$,
$p\, p \to K\, X$) for which experimental results
are or will soon be available.
\end{minipage}
\end{center}

Experimental data on SSA, $A_N$, for inclusive particle production
in polarized hadron-hadron scattering $A^\uparrow \,B\to C\,X$,
at moderately large $p_T$ and high energy, deeply challenge QCD
theoretical models.
It is well known that single spin effects in the hard,
elementary reactions are negligible;
several theoretical models incorporating higher-twist effects and
involving soft, non perturbative physics, have thus been proposed
in the last years.
They mainly differ in the physical mechanism responsible for the
observed effects, and/or for the stage of the full hadronic process
at which this mechanism acts (for a short review of both the
experimental and theoretical situation, see Ref. [1]
and references therein).
Here we limit ourselves to discuss in some details the approach we have
recently proposed in Ref. [1]. We will give a brief,
qualitative account of our approach, and will show how a simple
phenomenological implementation compares to available
experimental information and leads to predictions for processes
experimentally accessible in the near future.

We adopt a formalism based on QCD-improved parton models,
factorization theorems, and their extension to polarized processes,
allowing for transverse momentum ($\bfk_\perp$) effects in the partonic
distribution functions (DF) of the transversely polarized hadron.
In partonic language, analogously to the well known expression
of the unpolarized cross section for the process $AB\to C X$

\begin{equation}
d\sigma^{unp}=\frac{1}{2}(d\sigma^\uparrow+d\sigma^\downarrow) \sim
\sum_{a,b,c,d}\int\,dx_a\,dx_b\,\frac{1}{x_c}
f_{a/A}(x_a)f_{b/B}(x_b)\frac{d\hat\sigma}{d\hat t} D_{C/c}(x_c)
\label{dsun}
\end{equation}

\noindent
which enters the denominator of
$A_N=(d\sigma^\uparrow-d\sigma^\downarrow)/
(d\sigma^\uparrow+d\sigma^\downarrow)$, we get for the numerator
of $A_N$ (keeping only leading $\bfk_\perp$ effects)

\begin{equation}
d\sigma^\uparrow-d\sigma^\downarrow \sim
\sum_{a,b,c,d}\int\,dx_a\,dx_b\,\frac{1}{x_c}
\int d\bfk_{\perp_a} I^{a/A}_{+-}(x_a,\bfk_{\perp_a})
f_{b/B}(x_b)\frac{d\hat\sigma(\bfk_{\perp_a})}
{d\hat t} D_{C/c}(x_c)
\label{dspol}
\end{equation}

\noindent where $I^{a/A}_{+-}(x_a,\bfk_{\perp_a})$ takes into account
non perturbative, soft physics and plays, for SSA,
the same role plaid by the DF $f_{a/A}(x_a)$
in the unpolarized case.
Since $I^{a/A}_{+-}$ is an odd function of $\bfk_{\perp_a}$ [1]
we are forced to keep $\bfk_{\perp_a}$ effects also in the kinematics of the
partonic process. This makes the total effect a twist-3 one.

In order to apply this theoretical approach to practical situations,
we can take two possible attitudes: {\it i}\,) Try to find some
theoretical expression for  $I^{a/A}_{+-}$;
{\it ii}\,) Take $I^{a/A}_{+-}$ as a universal, phenomenological
non perturbative quantity, extract information on it from the available
data on SSA and use this information to produce
estimates of $A_N$ for other processes.

\begin{wrapfigure}{r}{8cm}
\epsfig{figure=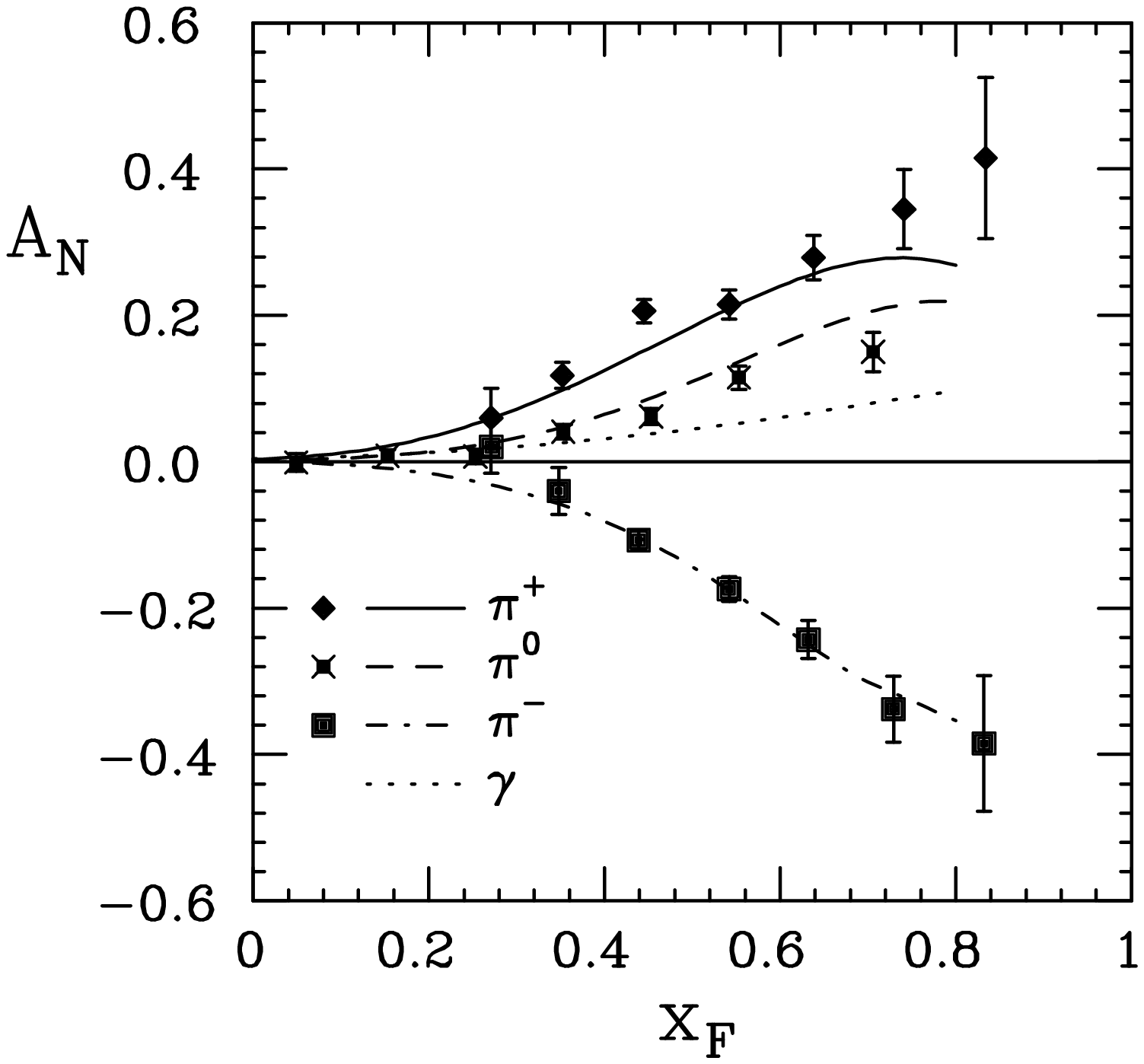,bbllx=50pt,bblly=200pt,bburx=500pt,%
bbury=620pt,width=8cm,height=8cm}
{\footnotesize {\bf Fig. 1}: Fit of the data on $A_N$ for $p^\uparrow p\to
\pi X$ [6]; predictions for $p^\uparrow p\to \gamma X$ are also
presented (see text)}
\end{wrapfigure}

In this paper we follow the second option [see Ref. [2]
for an attempt in the direction of point {\it i}\,)].
As a first approach, we assume that $I^{a/A}_{+-}$ is peaked around
an average value of $\bfk_{\perp_a}$, corresponding to the relevant
physical scale for transverse momentum effects in the polarized hadron.
Notice that this average value is in principle still dependent on $x_a$.
We fix this $x_a$ dependence following previous theoretical
work [3]. The remaining $x_a$ dependence of $I^{a/A}_{+-}$ not coming
{}from $\bfk_{\perp_a}$ is parametrized in the simple form
$N_a x_a^{\alpha_a} (1-x_a)^{\beta_a}$.
All other quantities appearing in Eqs.~(\ref{dsun}),(\ref{dspol})
are either theoretically or experimentally known:
{\it i}\,) for the unpolarized DF, $f_{a/A}(x)$,
we adopt here, as an example, the MRSG parametrization
[4]);
{\it ii}\,) for the fragmentation functions (FF) $D_{C/c}(x_c)$,
we take some set recently proposed for pions and kaons [5]).
{\it iii}\,) finally, the partonic cross sections
$d\hat\sigma/d\hat t$
are evaluated analytically at leading order in PQCD,
including modified kinematics due to $\bfk_{\perp_a}$
effects.

We then proceed fixing the
parameters ($N_a$, $\alpha_a$, $\beta_a$)
for  $I^{a/A}_{+-}$ (taking into account only valence
$u$, $d$ quark contributions, which dominate
in the large $x_F$ region where the effects
are observed) by a fit to the recent E704 results for
pion SSA [6], shown in Fig. 1. Since the data are for $p_T$ ranging from
0.7 to 2.0 GeV, we use a fixed $p_T$ value of 1.5 GeV.
%The quality of the fit is shown in Fig. 1.
The values of the parameters found are quite reasonable and agree with
qualitative expectations from simple $SU(6)$ arguments concerning the
polarization of $u$, $d$ quarks inside a transversely polarized proton.

Once we have fixed the free parameters of the model,
we can apply it to different cases.
For example, Fig. 2 shows how our model compares to the E704 results
for the $\bar p^\uparrow p \to \pi X$ process [6].

Another interesting process is $p^\uparrow (\bar p^\uparrow) p \to
\gamma X$, in that it can help to disentangle among
the different theoretical models proposed:
clearly all models involving effects in the final, fragmentation process
do not come into place here.
In Figs. 1,2 we  show, at the same c.m. energy as for the pion case,
but a higher $p_T$ value (2.8 GeV), our predictions for this process,
respectively for polarized proton and antiproton beams.

\begin{wrapfigure}{3}{8cm}
\epsfig{figure=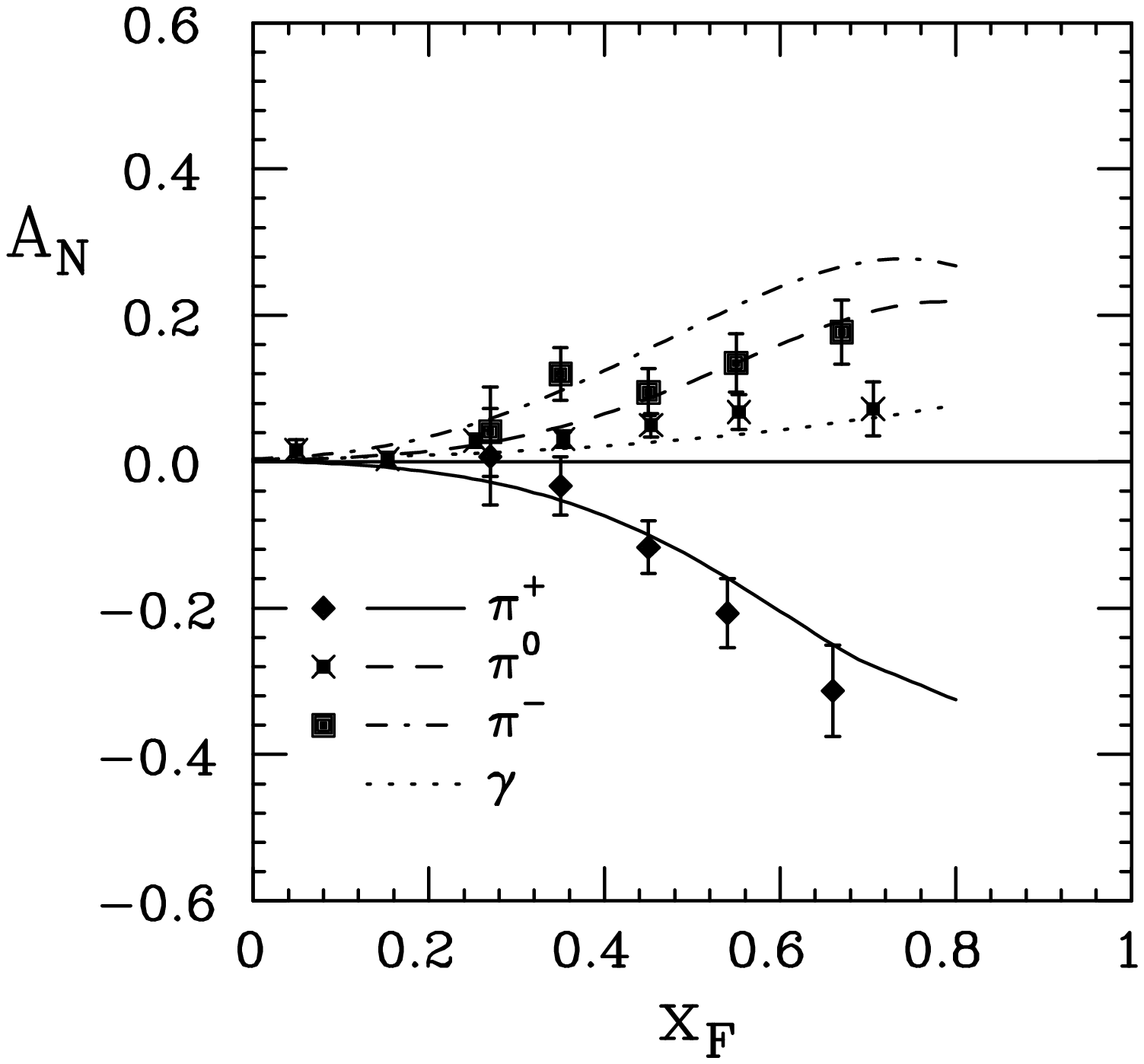,bbllx=50pt,bblly=200pt,bburx=500pt,%
bbury=620pt,width=8cm,height=8cm}
{\footnotesize {\bf Fig. 2}: $A_N$ for $\bar p^\uparrow p\to
\pi X$ (see [6] for data) and $\bar p^\uparrow p\to \gamma X$}
\end{wrapfigure}

Other interesting predictions involve the inclusive production of kaons.
However, at present little is known about kaon FF,
and this fact limit our capability of making predictions.
{}From a qualitative point of view, it is clear that if the ratio
between the FF for valence constituents (e.g., $D_{u/K^+}$)
and those for non-valence ones (e.g., $D_{s/K^+}$) is similar
to that in the pion, we expect that, for $p^\uparrow p\to K X$,
$A_N(K^+)\sim A_N(\pi^+)$; $A_N(K^0)\sim A_N(\pi^-)$,
$A_N(K^-,\bar K^0)\sim 0$. Preliminary results of our
model follow this behaviour, but show also a
certain sensibility to this ratio. More
refined FF are required to better clarify the kaon case. 

Let us finally mention that, in order to improve our
analysis, we need from the experimental side more
higher precision data, at higher energy and $p_T$; in particular, it
is essential to reach a better separation between $x_F$ and $p_T$
behaviours;
{}from the theoretical side, one of the main issues is certainly an
in-depth study of the interplay among the different
proposed theoretical mechanisms,
which may act simultaneously, adding up or canceling (even partially)
their effects.

\vspace{3pt}

\small
\begin{description}
\setlength{\parskip}{-5pt}
\item{[1]}
M.~Anselmino, M.~Boglione, and F.~Murgia, Phys. Lett. B362, 164 (1995).
\item{[2]}
M.~Anselmino, A.~Drago, and F.~Murgia, these proceedings.
\item{[3]}
J.D.~Jackson, G.G.~Ross, and R.G.~Roberts, Phys. Lett. B226, 159 (1989).
\item{[4]}
A.D.~Martin, W.J.~Stirling, and R.G.~Roberts, Phys. Rev. D50, 6734 (1994).
\item{[5]}
J.~Binnewies, B.A.~Kniehl, and G.~Kramer, Phys. Rev. D53, 3573 (1996),
and refs. therein.
\item{[6]}
E704 Collaboration, Phys. Lett. B264, 462 (1991);
report FERMILAB-pub-96/086-E, March 1996. 
\end{description}

\end{document}